# Two-hop Secure Communication Using an Untrusted Relay

Xiang He Aylin Yener
Wireless Communications and Networking Laboratory
Electrical Engineering Department
The Pennsylvania State University, University Park, PA 16802
*xxh119@psu.edu* *yener@ee.psu.edu*

**Abstract**

We consider a source-destination pair that can only communicate through an *untrusted* intermediate relay node. The intermediate node is willing to employ a designated relaying scheme to facilitate reliable communication between the source and the destination. Yet, the information it relays needs to be kept secret from it. In this two-hop communication scenario, where the use of the untrusted relay node is essential, we find that a positive secrecy rate is achievable. The center piece of the achievability scheme is the help provided by either the destination node with transmission capability, or an external “good samaritan” node. In either case, the helper performs cooperative jamming that confuses the eavesdropping relay and disables it from being able to decipher what it is relaying. We next derive an upper bound on the secrecy rate for this system. We observe that the gap between the upper bound and the achievable rate vanishes as the power of the relay node goes to infinity. Overall, the paper presents a case for intentional interference, i.e., cooperative jamming, as an enabler for secure communication.

This work was presented in part at the IEEE Globecom Conference, December 2008. This work is supported in part by the National Science Foundation with Grants CCR-0237727, CCF-051483, CNS-0716325, CNS-0721445 and the DARPA ITMANET Program with Grant W911NF-07-1-0028.

## I. Introduction

Information theoretic security was proposed by Shannon [1]. The idea of measuring secrecy using mutual information lends itself naturally to the investigation of how the channel can influence secrecy and further to the characterization of the fundamental limit of secure transmission rate. Wyner, in [2], defined the wiretap channel, and showed that secure communication from a transmitter to a "legitimate" receiver is possible when the signal received by the wiretapper (eavesdropper) is degraded with respect to that received by the legitimate receiver. Reference [3] identified the *secrecy capacity* of the general discrete memoryless wiretap channel. The secrecy capacity of the Gaussian wiretap channel is found in [4].

Recent progress in this area has extended classical information theory channel models to include secrecy constraints. Examples are the multiple access channel, the broadcast channel, the two-way channel, the three-node relay channel and the two-user interference channel [5]–[13]. These studies are beginning to lead to insights for designing secure wireless communication systems from the physical layer up. Prominent such examples include using multiple antennas to steer the transmitted signal away from an eavesdropper [14]–[16], transmitting with the intention of jamming the eavesdropper [8], [10], [17], and taking advantage of variations in channel state to provide secrecy [18]–[20].

The focus of this work is on a class of relay networks where the source and the destination have no direct link and thus can only communicate utilizing an intermediate relay node. This models the practical scenario where direct communication between the source and the destination is too "expensive" in terms of power consumption: Direct communication may be used to send some very low rate control packages, for example to initialize the communication, but it is infeasible to sustain a non-trivial reliable communication rate due to the power constraint.

In such a scenario, the source-destination pair *needs* the relay to communicate. On the other hand, more often than not, this relay node may be "untrusted" [11]. This does not mean the relay node is malicious, in fact quite the opposite, it may be part of the network and we will assume that it is willing to faithfully carry out the designated relaying scheme. The relay simply has a lower security clearance in the network and hence is not trusted with the confidential message it is relaying. Equivalently, we can assume the confidential message is one used for identification of the source node for authentication, which should never be revealed to a relay node in order not to be vulnerable to an impersonation attack. In all these cases, we must assume there is an eavesdropper co-located at the relay node when designing the system.

The "untrusted" relay model, or the eavesdropper being co-located with the relay node, was first studied in [9] for the general relay channel, with a rather pessimistic outlook, finding that for the degraded or the reversely degraded relay channel the relay node should not be deployed. More optimistic results for the relay channel with a co-located eavesdropper have been identified recently in references [11], [21], [22]. Specifically, it has been shown that the cooperation from the relay may, in fact, be essential to achieving non-zero secrecy rate [11], [21]. The model is later extended to the more symmetric case in [23], [24] where the relay also has a confidential message of its own, which must be kept secret from the destination.

All these models assume that a direct link between the source and the destination is present including our previous work [11]. In contrast, when there is no direct link, it is impossible for this network to convey a confidential message from the source to the destination while keeping it secret from the relay [9]. This is because the destination can only receive signals from the relay resulting in a physically degraded relay channel [25]. Therefore, the relay knows everything the destination knows regarding the confidential message, and the secrecy capacity is zero.

The differentiating feature of the model studied in this work from those described above including [11] is that the destination has transmission capability. This opens the possibility of the destination node to actively participate in ensuring the secrecy of the information it wants to obtain. In an effort to address a practical two-hop communication scenario, we shall consider each node to be half-duplex, which leads to a two-phase communication model. In addition, feedback to the source is not considered in the channel model. Interestingly, in this model, the transmission capability of the destination proves to be the *enabler* of secure communication. By recruiting the help of the destination to do "cooperative jamming", positive secrecy rate can be achieved that would not have been possible otherwise. We also remark that in case the transmission by the destination is not possible or desired, the help from an external cooperative jammer will do as well.

The idea of using a helpful jammer goes back to [17], [26], [27] and has since been used in many different models. Besides the multiple access, two-way [8] and relay wiretap channels [10], other recent results that use "cooperative jamming" as the part of the achievability scheme include [28]–[30]. In [28], a separate jammer is added to the classical Gaussian wiretap channel model. The jamming signal is revealed to the legitimate receiver via a wired link so that an advantage over the eavesdropper is gained. Reference [30] does not assume the wired connection, and employs a scheme tantamount to the two user multiple access channel with an external eavesdropper where one of the users perform cooperative jamming. Reference [29] considers the case where both the eavesdropper and the legitimate receiver observes a modulus $\Lambda$ channel and the destination carries out the jamming. We note that all these works deal with an external eavesdropper, in contrast to the focus of this work, which is an untrusted (but legitimate) node in the network.

In general, the optimality of recruiting a helpful jammer remains open as the converse results are limited. For the Gaussian case, the main difficulty is to find a upper bound for which the optimal input distribution can be found and evaluated. Doing so usually involves the introduction of genie information, as in the converse for the Gaussian wiretap channel [4], MIMO wiretap channel [14] and MAC wiretap channel [31]. The proofs then typically invoke the entropy power inequality, as in [4], or the generalized entropy power inequality, as in [32].

In this work, we derive a computable upper bound for the model in consideration by first introducing a second eavesdropper, an approach first used for a three-node relay channel in [11]. Next, after several steps of genie arguments, the channel is transformed into a wiretap channel with a helpful jammer, whose outer bound is then evaluated. The resulting bound is non-trivial in the sense that it is strictly tighter than the bound for the same channel without secrecy constraints. We also prove that it is tighter than an upper bound derived using the generalized entropy power inequality following a similar approach to [32], when the maximum sum received SNR at the relay is greater than $0$dB. We show that the gap between the bound and the achievable rates converges to zero when the power of the relay goes to $\infty$.

The paper is organized as follows. Section II presents the channel model and the two-phase protocol that utilizes cooperative jamming. In section III, we derive the achievable rates. Section IV presents our upper bound and compares with other known upper bounds. Section V presents the numerical results. Finally, Section VI presents the conclusion.

The following notation is used throughout this work: We use $H$ to denote the entropy, $h$ to denote the differential entropy, and $\varepsilon_k$ to denote any variable that goes to 0 when $n$ goes to $\infty$. We define $C(x) = \frac{1}{2}\log_2(1+x)$.

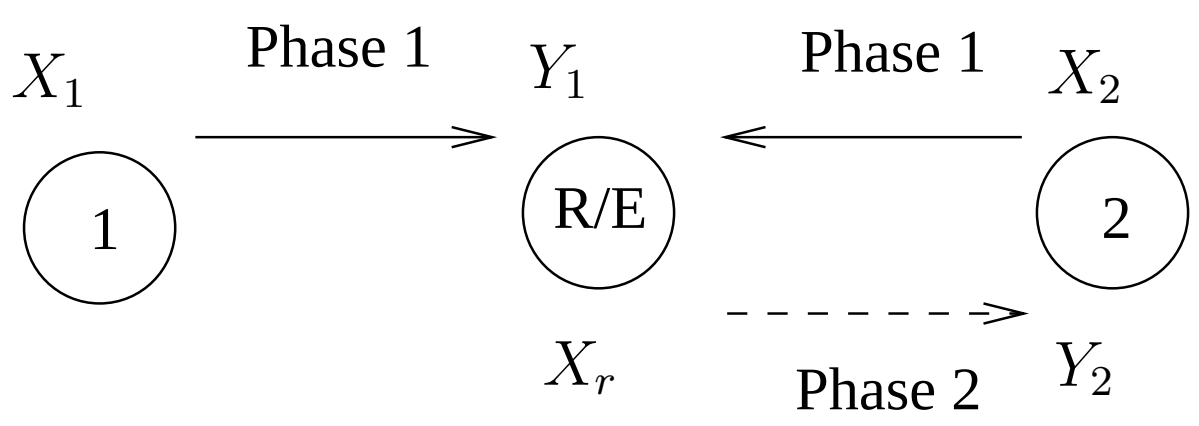

Fig. 1. Two-hop communication using an untrusted relay

## II. Channel Model

The system model is shown in Figure 1. We assume all nodes are half-duplex and the communication alternates between two phases, called phase one and phase two respectively. During phase one, shown with solid lines in Figure 1, the source transmits signal $X_1$. At the same time, the destination node transmits jamming signal $X_2$ in order to confuse the relay node. The signal received by the relay in phase one, $Y_1$, is given by

$$Y_1 = X_1 + X_2 + Z_1 \tag{1}$$

where $Z_1$ is a zero mean Gaussian random variable with unit variance. In an effort to reflect on the design of a practical system, we assume that the computation of $X_i$, $i = 1, 2$ does not rely on the signals received by node $i$ in the past.

In phase two, shown with dashed lines, the relay transmits signal $X_r$, which is computed from the local randomness at the relay, the signal transmitted and received by the relay in the past.

The signal received by the destination in phase two is denoted by $Y_2$, which is given below:

$$Y_2 = X_r + Z_2 \tag{2}$$

where $Z_2$ is a zero mean Gaussian random variable with unit variance.

The channel alternates between these two phases according to a random or deterministic schedule, which is generated by a global controller independently from the signals associated with the channel model. Hence here the term “schedule” is simply a finite number of channel uses which are either marked as phase one or phase two. We use $n$ to denote the number of channel uses marked as phase one, and $m$ to denote the number of channel uses marked as phase two. It should be noted that in general the $n$ channel uses of phase one are not consecutive. Neither are the $m$ channel uses of phase two. We assume the schedule is stable, in the sense that the following limit exists:

$$\alpha = \lim_{n+m\to\infty} \frac{n}{m+n} \tag{3}$$

For a given $\alpha$, we use $\{T(\alpha)\}$ to denote a sequence of schedules with increasing number of channel uses $n+m$ such that (3) holds. According to this definition, $\alpha$ becomes the limit of the time sharing factor of phase one in the schedule $T(\alpha)$ as $n+m\to\infty$.

When transmitting signals, the source, the destination, and the relay must satisfy certain power constraints. The average power constraints for the source, the jammer and the relay can be expressed as follows:

$$\frac{1}{N}\sum_{k=1}^{N} E\left[X_{i,k}^2\right] \leq \bar{P}_i, \quad i=1,2 \tag{4}$$

$$\frac{1}{N}\sum_{k=1}^{N} E\left[X_{r,k}^2\right] \leq \bar{P}_r \tag{5}$$

where

$$N = n+m \tag{6}$$

is the total number of channel uses.

For the purpose of completeness, we also introduce the notation $P_i, i=1,2$ to denote the average power of node $i$ during phase one. Since node $1$ and $2$ are only transmitting during phase one, $P_i$ and $\bar{P}_i$ are related as

$$P_i = \frac{\bar{P}_i}{\alpha}, \quad i=1,2 \tag{7}$$

Similarly, we use $P_r$ to denote the average power of the relay node during phase two. Since the relay node only transmits during the second phase, $P_r$ is related to $\bar{P}_r$ as follows:

$$P_r = \frac{\bar{P}_r}{1-\alpha} \tag{8}$$

After a number of phases, the destination node (node $2$) decodes a message $\hat{W}$ from the signals it transmitted during the periods of phase one and the signals it received during the periods of phase two. For reliable communication, $\hat{W}$ should equal the message $W$ from the source node with high probability. Hence we have the following requirement:

$$\lim_{n+m\to\infty} \Pr(W \neq \hat{W}) = 0 \tag{9}$$

The message $W$ must also be kept secret from the eavesdropper at the relay node, who can infer the value of $W$ based on the following knowledges available to it:

1) The local randomness at the relay, denoted by $A$.
2) The $n$ signals the relay transmitted during the periods of phase one, denoted by $Y_1^n$.
3) The $m$ signals the relay transmitted during the periods of phase two, denoted by $X_r^m$.

The information on $W$ that the eavesdropper can extract from these knowledges should be limited. Hence we have the following secrecy constraint.

$$\lim_{n+m\to\infty} \frac{1}{n+m} H\left(W\right) = \lim_{n+m\to\infty} \frac{1}{n+m} H\left(W|X_r^m, Y_1^n, A\right) \tag{10}$$

Since $W - \{X_r^m, Y_1^n\} - A$ is a Markov chain, we have

$$\lim_{n+m\to\infty} \frac{1}{n+m} H\left(W|X_r^m, Y_1^n, A\right) = \lim_{n+m\to\infty} \frac{1}{n+m} H\left(W|X_r^m, Y_1^n\right) \tag{11}$$

Therefore, the secrecy constraint can be expressed as

$$\lim_{n+m\to\infty} \frac{1}{n+m} H\left(W\right) = \lim_{n+m\to\infty} \frac{1}{n+m} H\left(W|X_r^m, Y_1^n\right) \tag{12}$$

For a given $\alpha$, and sequences of schedule $\{T(\alpha)\}$, the secrecy rate $R_e$ is defined as

$$R_e = \lim_{n+m\to\infty} \frac{1}{n+m} H\left(W\right) \tag{13}$$

such that (9) and (12) are fulfilled. When deriving achievable rate, we will focus on a specific sequence of schedules $\{T(\alpha)\}$, and maximize the secrecy rate over $\alpha$. When deriving the upper bound, we will consider all possible sequences of $\{T(\alpha)\}$.

*Remark 1:* Since the signals transmitted by node $1$ and $2$ do not depend on the signals they received in the past, $W \to Y_1^n \to X_r^m$ is a Markov chain. Therefore:

$$\lim_{n+m\to\infty} \frac{1}{n+m} H\left(W|X_r^m, Y_1^n\right) = \lim_{n+m\to\infty} \frac{1}{n+m} H\left(W|Y_1^n\right) \tag{14}$$

Hence, the following secrecy constraint can be used instead:

$$\lim_{n+m\to\infty} \frac{1}{n+m} H\left(W\right) = \lim_{n+m\to\infty} \frac{1}{n+m} H\left(W|Y_1^n\right) \tag{15}$$

*Remark 2:* We observe that in the system model shown in Figure 1, the destination, as the sender of the jamming signal during the periods of phase one, has perfect knowledge of this signal. This can be viewed as a special case of the model shown in Figure 2, where the destination only has a noisy copy of the jamming signal $Y_J = X_2 + Z_J$. If the jamming signal is corrupted by a noise sequence $Z_J$ that is independent of the noise sequences at the other receivers, then the secrecy capacity of the model in Figure 2 can not be larger than the secrecy capacity of the model in Figure 1. This is because giving this noise sequence to the destination as genie information would simply reveal the jamming signal $X_2$ to it. Therefore, any upper bound we derive for Figure 1 is also an upper bound for Figure 2. □

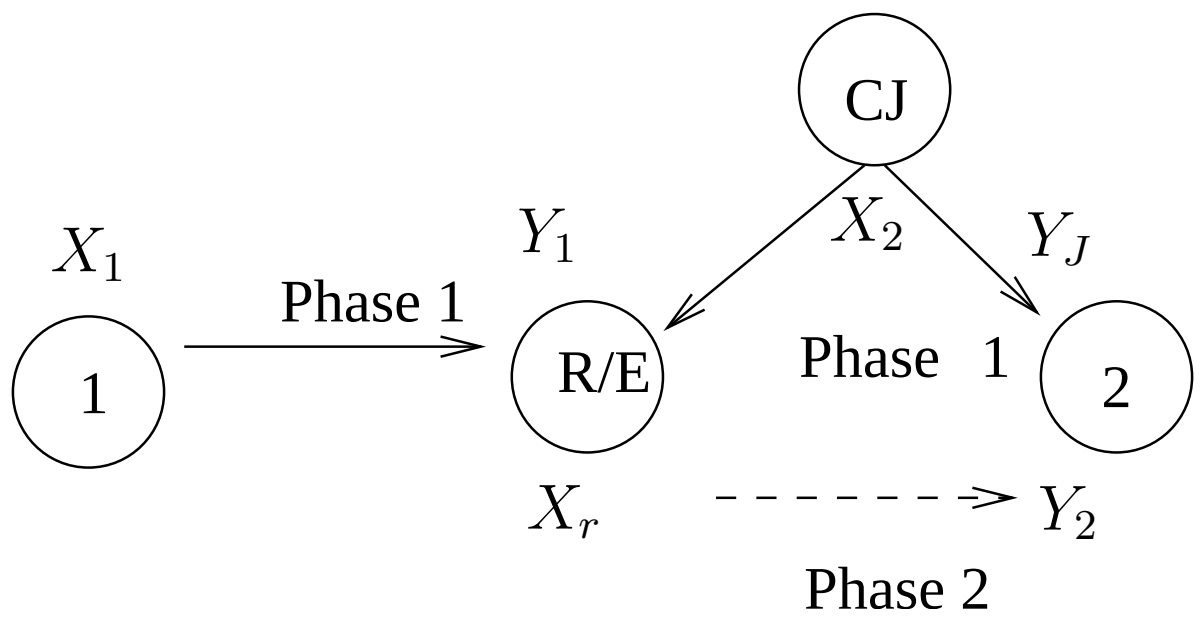

Fig. 2. Two-hop network with an external cooperative jammer, CJ

*Remark 3:* An apparent vulnerability of the described two phase protocol is that the destination may not be aware that the source has initiated its transmission. In this case, without the protection of the jamming signal from the destination, the message from the source would be revealed to the relay node and hence compromised. To prevent this from happening, proper initialization of the protocol is necessary. □

## III. Achievable Rate

In this section, we derive the achievable secrecy rate with the following sequence of deterministic periodic schedules:

The channel alternates between $n'$ channel uses for phase one and $m'$ channel uses for phase two, where $n'$ and $m'$ are two positive integers. The alternation takes $M$ times. Hence $n = n'M$ and $m = m'M$. For a given $\alpha$, the sequence of schedules is obtained by letting $M, n', m' \to \infty$ and

$$\lim_{n',m' \to \infty} \frac{n'}{m'+n'} = \alpha \tag{16}$$

With this sequence of schedules, we have the following theorem:

*Theorem 1:* The following secrecy rate is achievable for the model in Figure 1:

$$0 \leq R \leq \max_{0 \leq P_1' \leq \bar{P}_1/\alpha, 0<\alpha<1} \alpha \left[ C\left(\frac{P_1'}{(1+\sigma_c^2)}\right) - C\left(\frac{P_1'}{(1+P_2)}\right)\right]^+ \tag{17}$$

where $\sigma_c^2$ is the variance of the Gaussian quantization noise determined by:

$$\alpha C\left(\frac{P_1'+1}{\sigma_c^2}\right) = (1-\alpha) C\left(P_r\right) \tag{18}$$

where $P_2$ is defined in (7), $P_r$ is defined in (8).

*Proof:* The proof is given in Appendix A. ■

*Remark 4:* It can be seen from (17) that, for any fixed time sharing factor $\alpha$ the relay should always transmit at maximum power $P_r$. However, the optimal transmission power of the source may be less than $P_1$. This can be seen as follows: For a given jamming power $P_2$, the achievable rate is not a monotonically increasing function of $P_1'$. This is because, if $P_1' \to 0$ or $P_1' \to \infty$, $R_e \to 0$, indicating that even if the source power budget is $\infty$, the optimal transmission power is actually finite. Let this value be $P_1^*$. $P_1^*$ may or may not fall into the interval $[0, P_1]$, which is the range of power consumption allowed for phase one. If it does, then the source should transmit with power $P_1^*$ rather than $P_1$. If not, then the corresponding optimal value needs to be determined. □

*Remark 5:* If the power constraint of the relay $\bar{P}_r \to \infty$, then $\sigma_c^2 \to 0$, $\alpha \to 1$. The achievable rate converges to

$$C(\bar{P}_1) - C(\frac{\bar{P}_1}{1+\bar{P}_2}) \tag{19}$$

□

## IV. Upper bound

In this section, we derive an upper bound for the secrecy rate.

We first need to determine the optimal schedule. It turns out that it is easy to find: We simply let the first $n$ channel uses be phase one, and the remaining $m$ channel uses be phase two. The optimality of this schedule can be proved as following:

Suppose a different schedule is used. Since the signals received in the past are not used for encoding purposes at node $1$ and $2$, we can always move the channel uses of phase one to the front without affecting the signals transmitted by these two nodes. On the other hand, we notice that the relay can only use signals received in the past to compute its transmission signals.

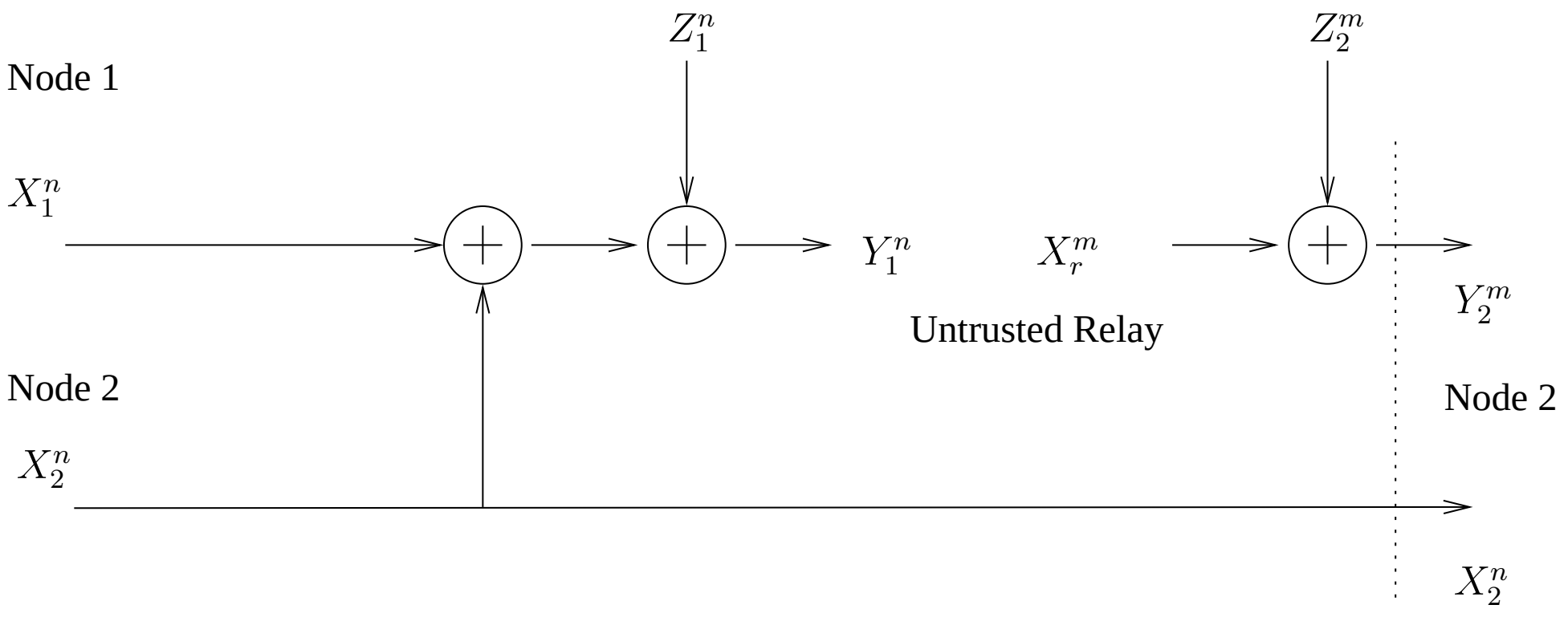

Fig. 3. Equivalent Channel Model for Deriving the Upper Bound

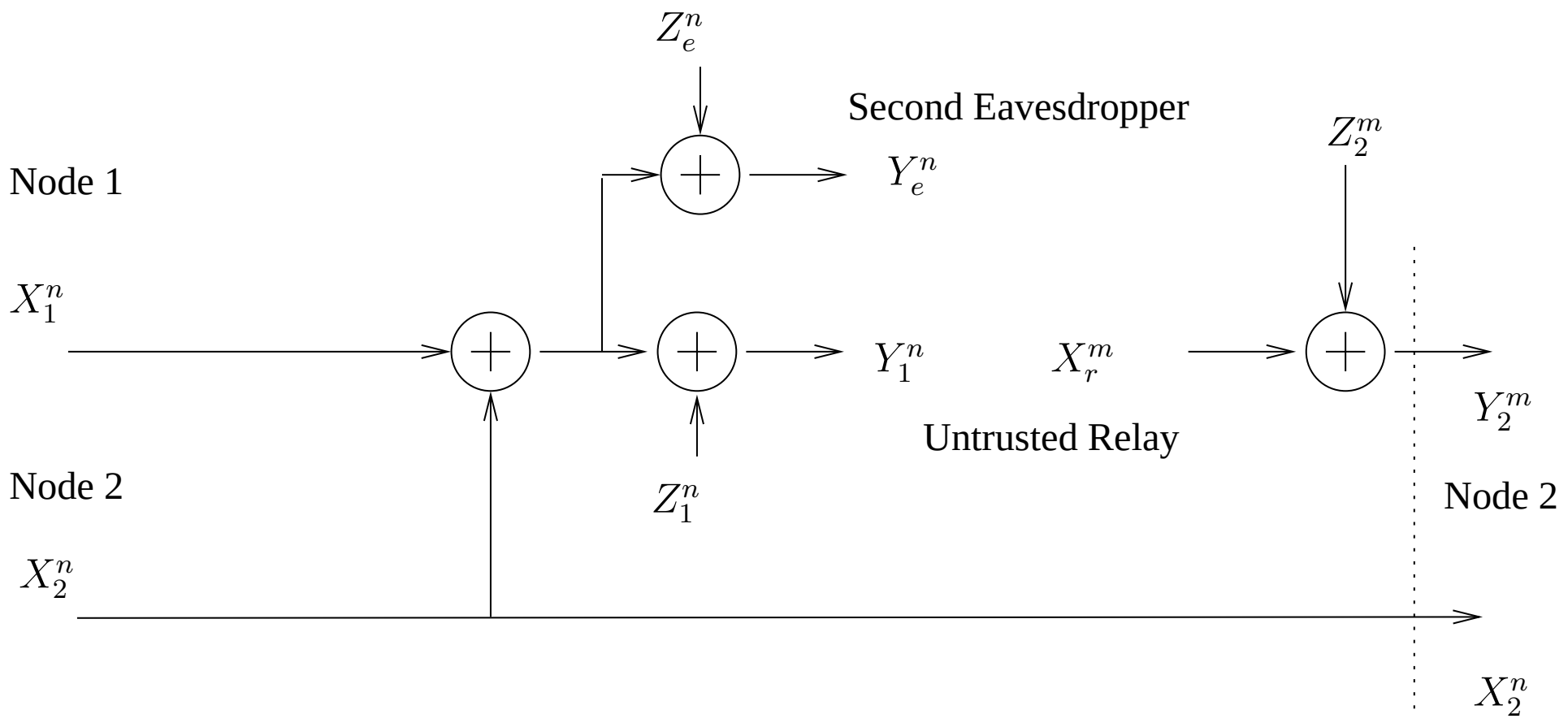

Fig. 4. Two-eavesdropper channel

However, during phase one, the relay only receives signals. Since moving phase one ahead only means the relay could receive signals sooner, doing so will not limit the capability of the relay to calculate its transmitted signals. Consequently, we observe that no matter what schedule is used to achieve a secrecy rate, we can always modify this schedule such that all channel uses of phase one are ahead of those of phase two and still achieve the same secrecy rate. Hence in the following we only consider the optimal schedule.

We also observe we can transform the channel into the one shown in Figure 3. The jammer and the receiver are now drawn separately, since the jammer does not use the signal received in the past to compute the jamming signal. Note that Figure 3 is similar to Figure 12 used in the achievability proof except that the dimension of the signals is changed from $m', n'$ to $m, n$.

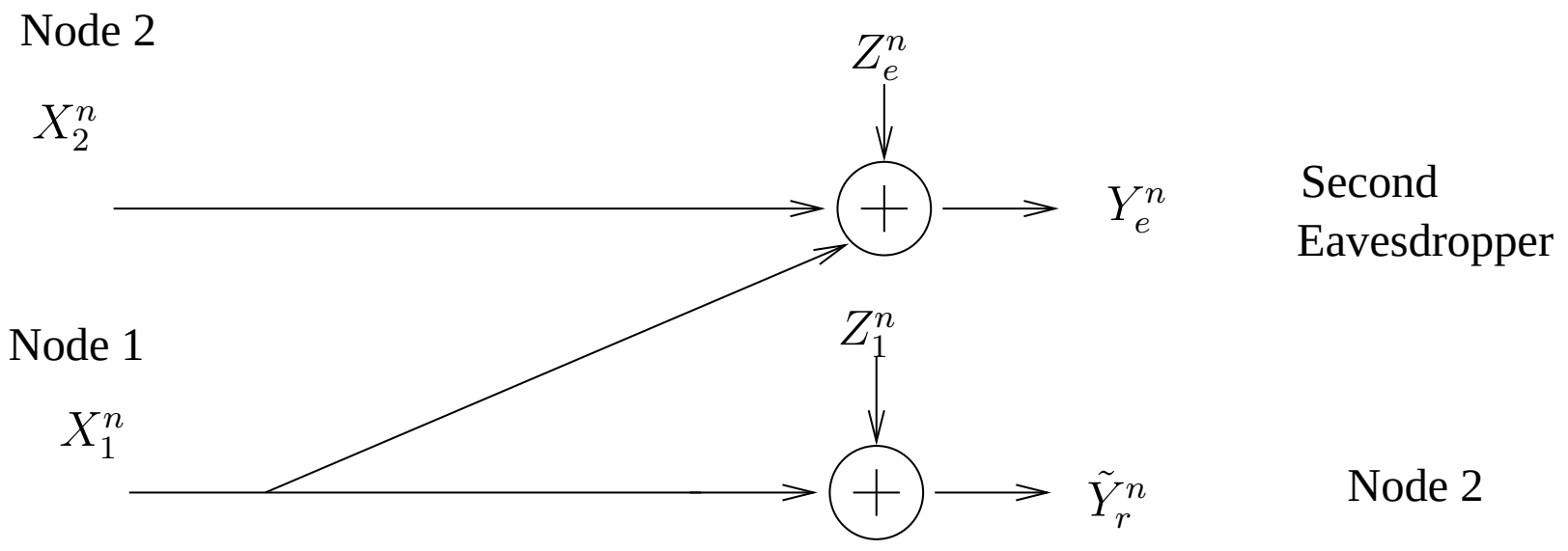

Fig. 5. Channel model after transformation

We next leverage a technique first used in [11], [22] to derive the upper bound. Specifically, the upper bound is obtained via the following transformations:

1) First, we add a second eavesdropper to the channel, as shown by Figure 4. Its received signal is denoted by $Y_e$ and over $n$ channel uses $Y_e^n$ is given by:

$$Y_e^n = X_1^n + X_2^n + Z_e^n \tag{20}$$

Here $Z_e^n$ is a Gaussian noise with the same distribution as $Z_1^n$. $Z_e^n$ can be arbitrarily correlated with $Z_1^n$. Since

$$Y_1^n = X_1^n + X_2^n + Z_1^n \tag{21}$$

We have

$$\Pr(W, Y_e^n) = \Pr(W, Y_1^n) \tag{22}$$

Therefore

$$H\left(W|Y_e^n\right) = H\left(W|Y_1^n\right) \tag{23}$$

From (15), this means

$$\lim_{n+m\to\infty} \frac{1}{m+n} H(W) = \lim_{n+m\to\infty} \frac{1}{m+n} H(W|Y_e^n) \tag{24}$$

Hence the message $W$ is kept secret from the second eavesdropper. This means, for a given coding scheme that achieves secrecy rate in Figure 3, the same secrecy rate is achievable with the introduction of this additional eavesdropper.

2) Next, we remove the first eavesdropper at the relay. Doing so will not decrease secrecy rate either, since we have one less secrecy constraint.

From (24), the secrecy rate can be upper bounded via $H(W|Y_e^n)$. To do that, we provide the signal $X_r^m$ to the destination by a genie. Similarly, the signal $X_2^n$ is revealed to both the relay and the destination. $H(W|Y_e^n)$ is then bounded by:

$$
\begin{aligned}
&H\left(W|Y_e^n\right)\\
\leq &H\left(W|Y_e^n\right) - H\left(W|X_r^m Y_2^m X_2^n\right) + n\varepsilon && (25)\\
= &H\left(W|Y_e^n\right) - H\left(W|X_r^m X_2^n\right) + n\varepsilon && (26)\\
\leq &H\left(W|Y_e^n\right) - H\left(W|Y_1^n X_r^m X_2^n\right) + n\varepsilon && (27)\\
= &H\left(W|Y_e^n\right) - H\left(W|Y_1^n X_2^n\right) + n\varepsilon && (28)\\
= &H\left(W|Y_e^n\right) - H\left(W|X_1^n + Z_1^n\right) + n\varepsilon && (29)\\
\leq &H\left(W|Y_e^n\right) - H\left(W|Y_e^n, X_1^n + Z_1^n\right) + n\varepsilon && (30)
\end{aligned}
$$

The genie information $X_r^m$ causes the signal $Y_2^m$ to be useless to the relay, as shown by (25)-(26). (28) is due to the fact that once the signal received by the relay $Y_1^n$ is given, the signal transmitted by the relay $X_r^m$, which is computed from $Y_1^n$, is independent from the jamming signal $X_2^n$ and the confidential message $W$. Finally, revealing the genie information $X_2^n$ to the relay and the destination essentially removes the influence of the jamming signal from the relay link, as shown by (28)-(30). These are essentially a consequence of the link noises being independent. The resulting channel is equivalent to the one shown in Figure 5, and can be viewed as a special case of the channel in [8], [33]. Similar techniques to those in [31], [33] can be used here to bound the secrecy rate. Let $\tilde{Y}_r^n = X_1^n + Z_1^n$. Then (30) becomes:

$$
\begin{aligned}
&H\left(W|Y_e^n\right) - H\left(W|Y_e^n \tilde{Y}_r^n\right) && (31)\\
= &I\left(W; \tilde{Y}_r^n | Y_e^n\right) && (32)\\
\leq &I\left(W X_1^n; \tilde{Y}_r^n | Y_e^n\right) && (33)\\
= &I\left(X_1^n; \tilde{Y}_r^n | Y_e^n\right) && (34)\\
= &h\left(\tilde{Y}_r^n | Y_e^n\right) - h\left(Z_1^n | X_2^n + Z_e^n\right) && (35)\\
\leq &h\left(\tilde{Y}_r^n | Y_e^n\right) - h\left(Z_1^n | X_2^n + Z_e^n, X_2^n\right) && (36)\\
= &h\left(\tilde{Y}_r^n | Y_e^n\right) - h\left(Z_1^n | Z_e^n\right) && (37)
\end{aligned}
$$

Here (34) follows from the fact that $X_1^n$ determines $W$. The first term in (37) is maximized when $X_1^n$ and $X_2^n$ are i.i.d. Gaussian sequences [14]. Let the variance of each component of $X_i^n$ be $P_i = \bar{P}_i/\alpha$, $i = 1, 2$. Let $\rho$ be the correlation factor between $Z_1$ and $Z_e$. Then (37) is equal to

$$\frac{n}{2}\log_2 \frac{(P_1+1)(P_1+P_2+1)-(P_1+\rho)^2}{(P_1+P_2+1)(1-\rho^2)} \tag{38}$$

It can be verified that, for any fixed $\rho$, equation (37) is an increasing function of $P_1$ and $P_2$. Therefore, the upper bound is maximized with maximum average power. Equation (38) can then be tightened by minimizing it over $\rho$. The optimal $\rho$ is given below:

$$\frac{2P_1 + P_1P_2 + P_2 - \sqrt{A}}{2P_1} \tag{39}$$

where

$$A = 4P_2P_1^2 + 4P_2P_1 + P_2^2P_1^2 + 2P_2^2P_1 + P_2^2 \tag{40}$$

As a result, we have the following theorem:

*Theorem 2:* The secrecy rate of the channel in Figure 12 is upper bounded by

$$\max_{0<\alpha<1} \min\left\{\frac{\alpha}{2}\log_2 \frac{(P_1+1)(P_1+P_2+1)-(P_1+\rho)^2}{(P_1+P_2+1)(1-\rho^2)}, (1-\alpha)C(P_r)\right\} \tag{41}$$

where $\rho$ is given by (39). $P_1 = \bar{P}_1/\alpha$, $P_2 = \bar{P}_2/\alpha$, and $P_r = \bar{P}_r/(1-\alpha)$ are the average power constraints of node $1, 2$ and the relay for the time sharing factor $\alpha$.

*Remark 6:* If we further fix $\bar{P}_2$ , and let $\bar{P}_r, \bar{P}_1 \to \infty$, then $\alpha \to 1$. $\rho$ converges to $\bar{\rho}$ given by:

$$\bar{\rho} = 1 + \bar{P}_2/2 - \sqrt{\bar{P}_2 + \bar{P}_2^2/4} \tag{42}$$

The difference of the upper bound and the achievable rate converges to

$$C\left(\frac{\bar{P}_2 + (\bar{\rho}-1)^2}{1-\bar{\rho}^2}\right) - C\left(\bar{P}_2\right) \tag{43}$$

We observe that the difference is only a function of $\bar{P}_2$. By comparison, the gap between the achievable rate and the trivial upper bound $C(\bar{P}_1)$ is $C(\frac{\bar{P}_1}{1+\bar{P}_2})$, which is unbounded. □

*Remark 7:* If we instead fix $\bar{P}_2 = \beta\bar{P}_1$, and let $\bar{P}_r \to \infty$, then $\alpha \to 1$. The achievable rate converges to (19). In this case, if we further let $\bar{P}_1 \to \infty$, the upper bound given by (41) converges to

$$C(\bar{P}_1) - C(1/\beta) \tag{44}$$

Comparing it with (19), we observe the difference of the upper bound and the achievable rate converges to $0$. Hence, in this case, our upper bound is asymptotically tight. □

*Remark 8:* The first term in the bound (41) is strictly smaller than the trivial bound $\alpha C(P_1)$ obtained by removing the secrecy constraints. To show that, simply let $\rho = 0$. (41) becomes

$$\alpha C\left(P_1\right) + \frac{\alpha}{2}\log_2 \frac{1+\frac{P_1}{(P_1+1)(P_2+1)}}{1+\frac{P_1}{(P_2+1)}} \tag{45}$$

The second term in (45) is always negative. □

### A. Comparison with the bound derived with generalized Entropy Power Inequality

Recently the generalized entropy power inequality [34] was used to derive a computable upper bound for the Gaussian multiple access channel with secrecy constraints [32]. Here the same technique is applicable and another computable upper bound for the model in Figure 1 can be derived. It is of interest to know which bound is tighter. Next, we prove that as long as $P_1 + P_2 > 1$, this upper bound is always looser than the bound given by (38)-(39).

First, we briefly describe the derivation of the bound based on the approach in [32]:

$$\begin{aligned}
&H\left(W|Y_1^n\right) \\
\overset{(a)}{=}&H\left(W|Y_1^n\right) - H\left(W|Y_1^n X_2^n\right) + n\varepsilon && (46)\\
=&I\left(W;X_2^n|Y_1^n\right) + n\varepsilon && (47)\\
\le&I\left(WX_1^n;X_2^n|Y_1^n\right) + n\varepsilon && (48)\\
=&I\left(X_1^n;X_2^n|Y_1^n\right) + n\varepsilon && (49)
\end{aligned}$$

Here step $(a)$ follows from Fano's inequality. (49) can be written as:

$$\begin{aligned}
&I\left(X_1^n;Y_1^n|X_2^n\right) + I\left(X_1^n;X_2^n\right) - I\left(X_1^n;Y_1^n\right) + n\varepsilon && (50)\\
\overset{(b)}{=}&I\left(X_1^n;Y_1^n|X_2^n\right) - I\left(X_1^n;Y_1^n\right) + n\varepsilon && (51)\\
=&h\left(X_1^n+Z_1^n\right) + h\left(X_2^n+Z_1^n\right) - h\left(Z_1^n\right) - h\left(X_1^n+X_2^n+Z_1^n\right) + n\varepsilon && (52)
\end{aligned}$$

Step $(b)$ follows from $X_1^n, X_2^n$ being independent.

Next, like [32, (76)], we invoke the inequality from [34] and obtain

$$2^{\frac{2}{n}h\left(X_1^n+X_2^n+Z_1^n\right)} \ge \frac{2^{\frac{2}{n}h\left(X_1^n+Z_1^n\right)} + 2^{\frac{2}{n}h\left(X_2^n+Z_1^n\right)}}{2} \tag{53}$$

Hence (52) can be upper bounded with

$$h\left(X_1^n+Z_1^n\right)+h\left(X_2^n+Z_1^n\right)-\frac{n}{2}\log_2\left(2^{\frac{2}{n}h\left(X_1^n+Z_1^n\right)}+2^{\frac{2}{n}h\left(X_2^n+Z_1^n\right)}\right)+\frac{n}{2}\log_2(2) \tag{54}$$

This expression is maximized when $X_1^n, X_2^n$ are chosen to be i.i.d. Gaussian sequences. Dividing by the total number of channel uses $n+m$, the final expression of the upper bound is given by

$$\frac{\alpha}{2}\log_2\left(\frac{2(P_1+1)\left(P_2+1\right)}{P_1+P_2+2}\right) \tag{55}$$

*Remark 9:* Note that (55) is also tighter than the bound $\alpha C(P_1)$ when $P_1 > P_2$. Hence it is a nontrivial bound when $P_1 > P_2$. □

*Remark 10:* When $\bar{P}_r \to \infty$, then $\alpha \to 1$, $P_i \to \bar{P}_i, i = 1, 2$. Comparing (19) with (55), the gap between the achievable rates and the bound given by (55) is

$$\frac{1}{2}\log_2\left(1+\frac{\bar{P}_1+\bar{P}_2}{2+\bar{P}_1+\bar{P}_2}\right) \tag{56}$$

which is smaller than $0.5$ bit/channel use. □

We next show that for any given $\alpha$, if $P_1+P_2 > 1$, (55) is always bigger than the first term in (41). We omit the time sharing factor $\alpha$ in the front since they are present in both expressions. Then we pick $\rho$ such that

$$1-\rho^2=\frac{P_1+P_2+2}{2\left(P_1+P_2+1\right)} \tag{57}$$

Note that this is a valid choice for $\rho$ since the right hand side is within the interval $(0, 1)$. Then equation (41), after canceling $\alpha$ in the front, becomes

$$\frac{1}{2}\log_2\frac{2\left(\left(P_1+1\right)\left(P_1+P_2+1\right)-\left(P_1+\rho\right)^2\right)}{P_1+P_2+2} \tag{58}$$

Hence we only need to verify that (55) is greater than (58) when $P_1+P_2 > 1$. This is equivalent to verifying

$$\left(P_1+1\right)\left(P_2+1\right) > \left(P_1+1\right)\left(P_1+P_2+1\right)-\left(P_1+\rho\right)^2 \tag{59}$$

Or $(2\rho-1)\,P_1+\rho^2 > 0$. A sufficient condition for this to hold is to require $2\rho-1 > 0$. Substitute (57) into this requirement we get $P_1+P_2 > 1$.

*Remark 11:* Since the gap between the achievable rates and the bound given by (55) is bounded by $0.5$ bit/channel use when $\bar{P}_r \to \infty$, the gap between the achievable rates and the bound given by (41) is also bounded by $0.5$ bit/channel use when $\bar{P}_r \to \infty$ and $\bar{P}_1 + \bar{P}_2 > 1$. Note that since when $\bar{P}_r \to \infty$ we have $\alpha \to 1$, the condition $P_1 + P_2 > 1$ is equivalent to $\bar{P}_1 + \bar{P}_2 > 1$. □

*Remark 12:* For the case that $P_1 + P_2 < 1$, it is not clear between (55) and (41) which bound is tighter. However, for these cases, the secrecy capacity is so small that the bounds are of no consequence. □

## V. Numerical Results

| | Relay's Power | Jammer's Power | $\alpha$ |
|---|---|---|---|
| Fig. 6 | $\infty$ | Proportional | optimal |
| Fig. 7 | $\infty$ | Fixed | optimal |
| Fig. 8 | Limited | Proportional | 0.5 |
| Fig. 9 | Limited | Fixed | 0.5 |
| Fig. 10 | Limited | Proportional | optimal |
| Fig. 11 | Limited | Fixed | optimal |

TABLE I
SCENARIOS CONSIDERED IN THE NUMERICAL RESULTS

Shown in Table I are the six cases of interest, corresponding to different power budgets of the relay and the jammer and whether time sharing factor $\alpha$ is fixed. We included the cases with fixed time sharing factor because in a real system, for simplicity the time sharing factor may not be dynamically adjusted according to power budgets. The numerical result of each case is shown in the figures listed in the table. We stress that, though not explicitly considered in the numerical results, for the more general case where the cooperative jammer is external, as shown in Figure 2, the upper bound still holds, but the gap between the upper bound and achievable rates would be wider.

Figures 6 and 7 demonstrate the asymptotic behavior described in Remark 6 when the power of relay goes to $\infty$. Note that in this case the optimal time sharing factor $\alpha$ converges to 1. Figures 8 and 9 demonstrate the case where the power the relay is finite, and the time sharing factor $\alpha$ is fixed. In all four cases, we observe the upper bound is close to the achievable rate when relay's power is larger than the power of the source and the jammer. In this region, typically, the achievable rate increases linearly with the source SNR. In Figure 6, the gap between the upper bound and achievable rate goes to zero as $P_1 \to \infty$. In Figure 7, the upper bound almost coincides with the achievable rate. The gap, given by (43), equals $9.98 \times 10^{-4}$ bits/channel use.

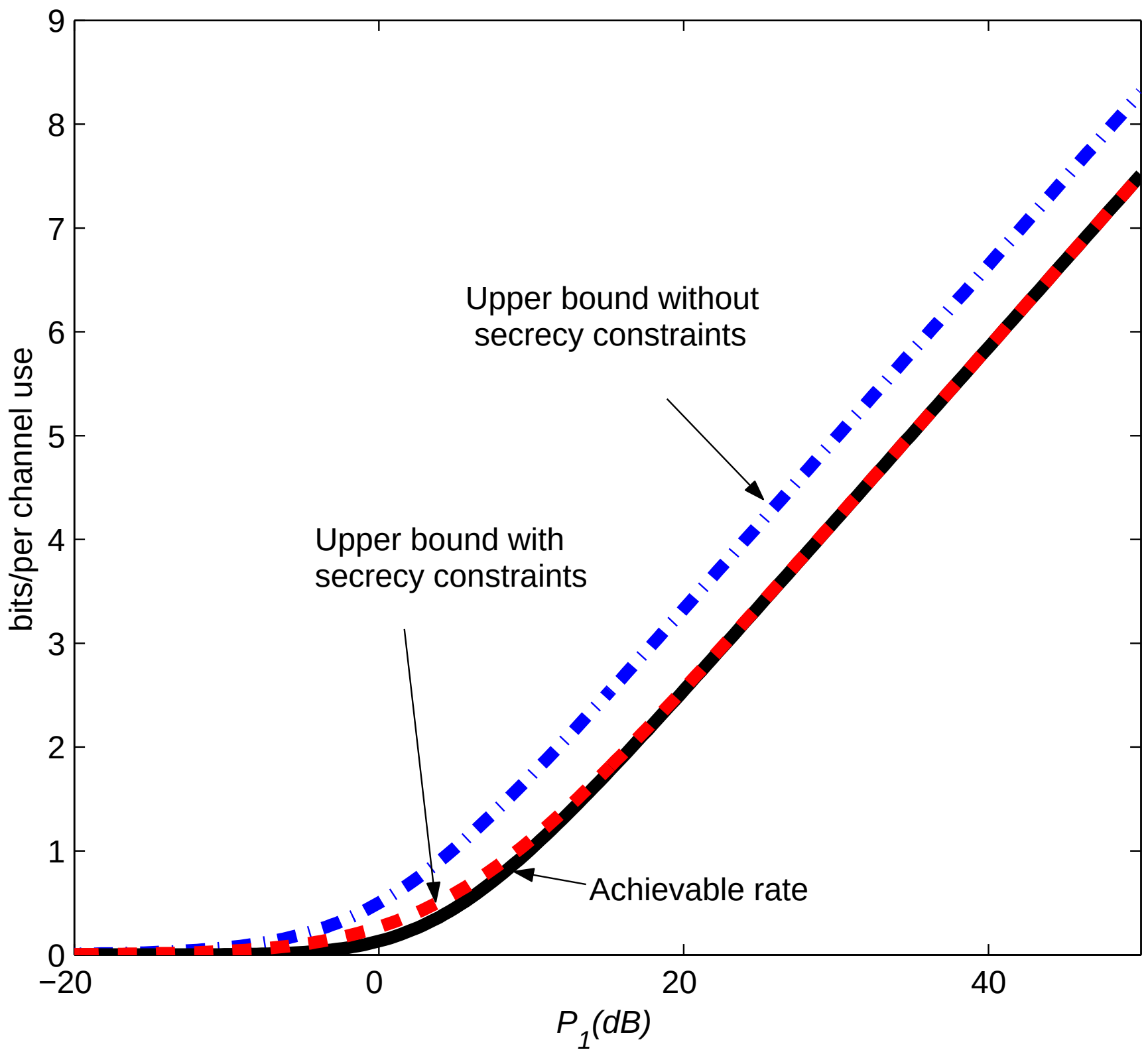

Fig. 6. Secrecy Rate, $P_r \to \infty$, $P_2 = 0.5P_1$, optimal $\alpha$

Also shown in each figure is the cut-set bound without secrecy constraints. The improvement provided by the new bounds depends on the power budget. In general, the improvement is small if the power of the jammer is large. Note that since we have normalized all channel gains and included them into the power constraint, the power budget difference can be considered a consequence of the difference in link gains.

Figure 9 also illustrates the power control problem described in Remark 4. Without power control at the source node, the achievable rate will eventually decrease to zero. Note that this behavior crystallizes only when the relay's power is limited.

Finally, in Figure 10 and Figure 11, we compare the achievable rates and the upper bound when each are maximized over the time sharing factor $\alpha$. The gap between the upper bound and the achievable rate is now wider because the second term in the upper bound (41) is the same as the upper bound without secrecy constraint. The role played by the second term (41) becomes significant when the bound is optimized over the time sharing factor, which as pointed out in [35], has a tendency to balance the two terms in the bound (41). However, as shown in these figures, compared to the upper bound without secrecy constraints, the new bound still offers significant improvement.

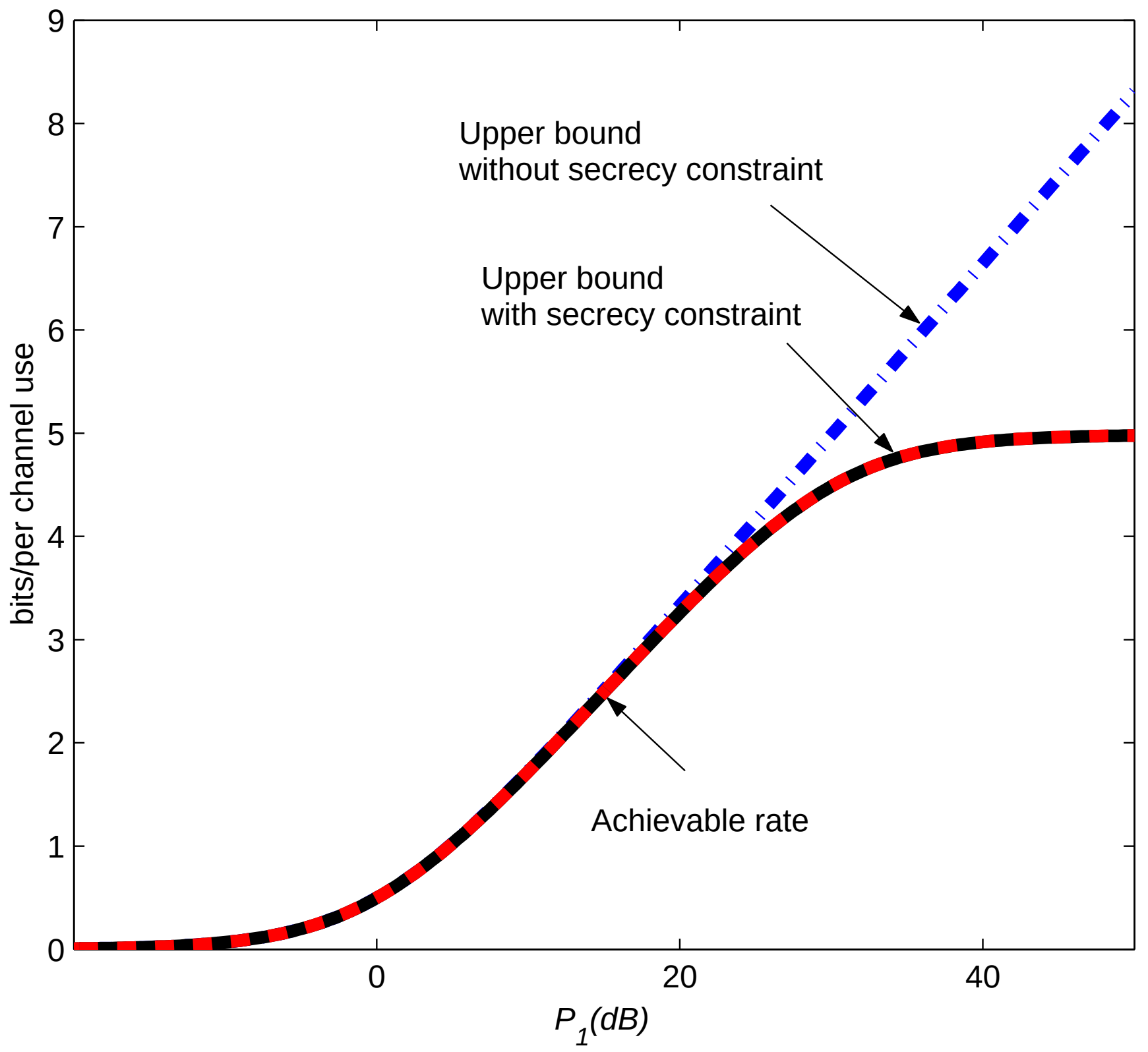

Fig. 7. Secrecy Rate, $P_r \to \infty$, $P_2 = 30$dB, optimal $\alpha$

## VI. Conclusion

In this paper, we have considered a relay network without a direct link, where relaying is essential for the source and the destination to communicate despite the fact that the relay node is untrusted. Imposing secrecy constraints at the relay node, contrary to the previous work, we have shown that a nonzero secrecy rate is indeed achievable. This is accomplished by enlisting the help of the destination (or another dedicated node) who transmits to jam the relay, and uses the jamming signal as side information. We have derived an upper bound for the secrecy rate with the assumption that no feedback is used for encoding at the source or destination. The new upper bound is strictly tighter than the upper bound without secrecy constraints. We have also proved that it is tighter than an upper bound derived from generalized entropy power inequality when the maximum sum received SNR at the relay is greater than $0$dB. The gap between the bound and the achievable rates converges to $0$ when the power of the transmitter, the relay and the jammer goes to $\infty$. Numerical results show that our upper bound is in general close to the achievable rate, and is indistinguishable from it for a fixed time sharing factor with a relay whose power is in abundance.

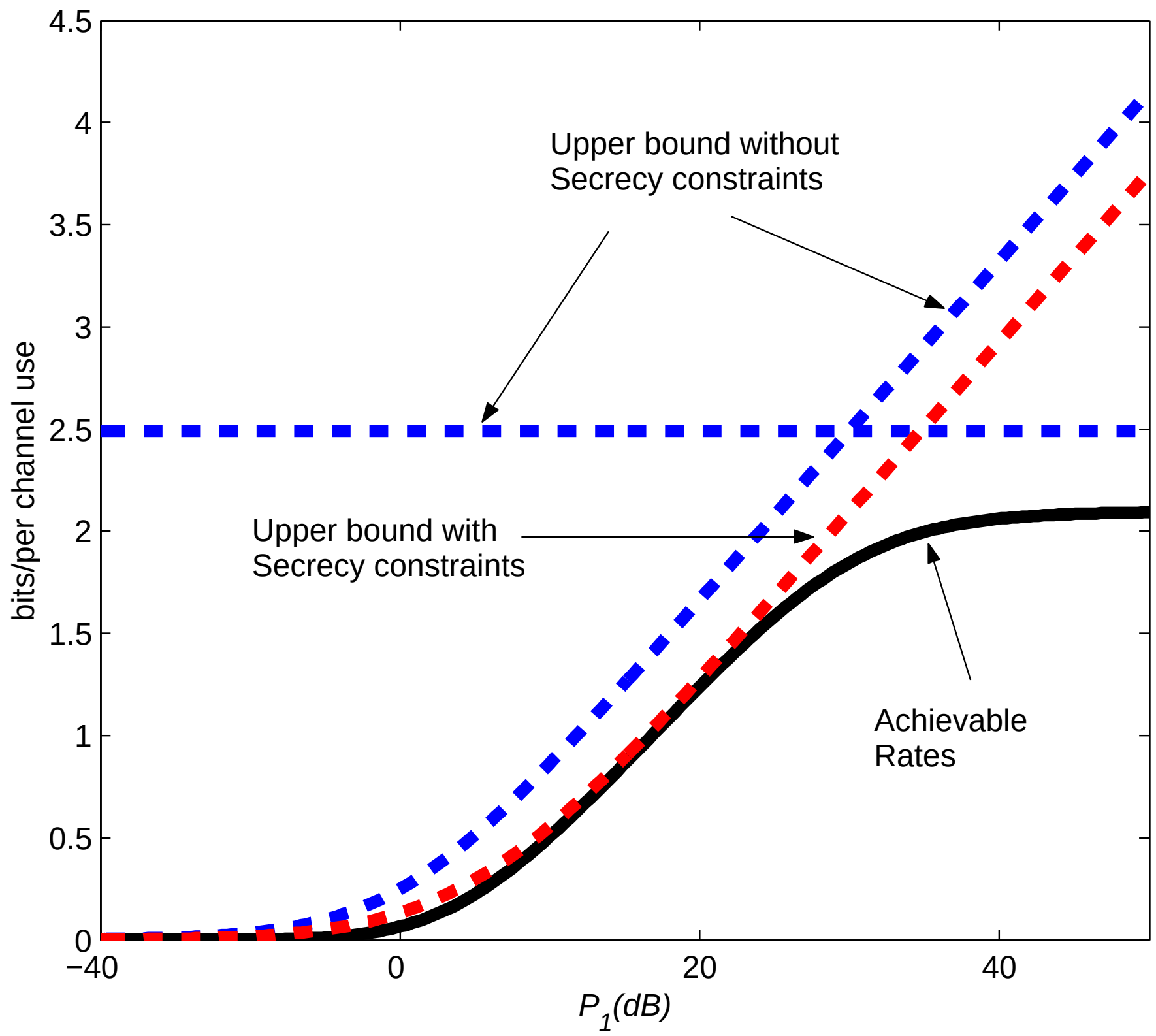

Fig. 8. Secrecy Rate, $P_r = 30$dB, $P_2 = 0.5P_1$, $\alpha = 0.5$

In this work, we considered the case where the source or the jammer does not make use of the relay transmission for encoding purposes. An upper bound for the secrecy rate when feedback is used is recently found in [36]. A gap exists between the upper bound and the achievable rate in [36], which is bounded by $0.5$bit per channel use but does not vanish when the power of the transmitter, the relay and the jammer goes to $\infty$. By comparison, the bound presented in this work is asymptotically tighter in this case.

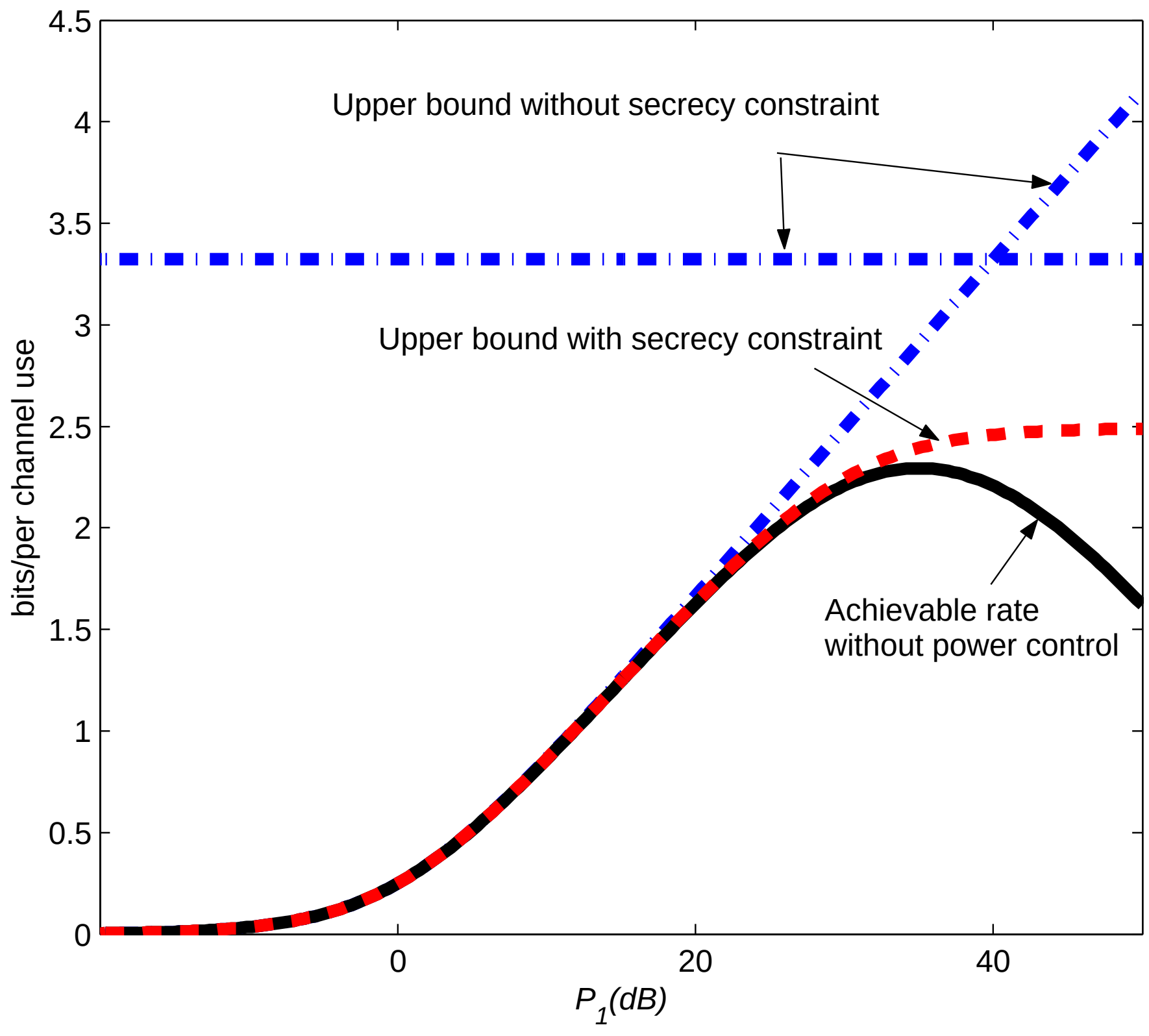

Fig. 9. Secrecy Rate, $P_r = 30$dB, $P_2 = 40$dB, $\alpha = 0.5$

We conclude by reiterating that our findings in this paper presents cooperative jamming as an enabler for secrecy from an *internal* eavesdropper, and motivates further investigation of such cooperation ideas in more general settings including those in larger networks. We also comment whether and when cooperative jamming actually yields the secrecy capacity (region) for various multiuser channels remain open problems in information theory.

## Appendix A
## Proof of Theorem 1

We first introduce several supporting results used in proving Theorem 1.

In reference [11], [21], we presented the following achievable secrecy rate for a general relay channel:

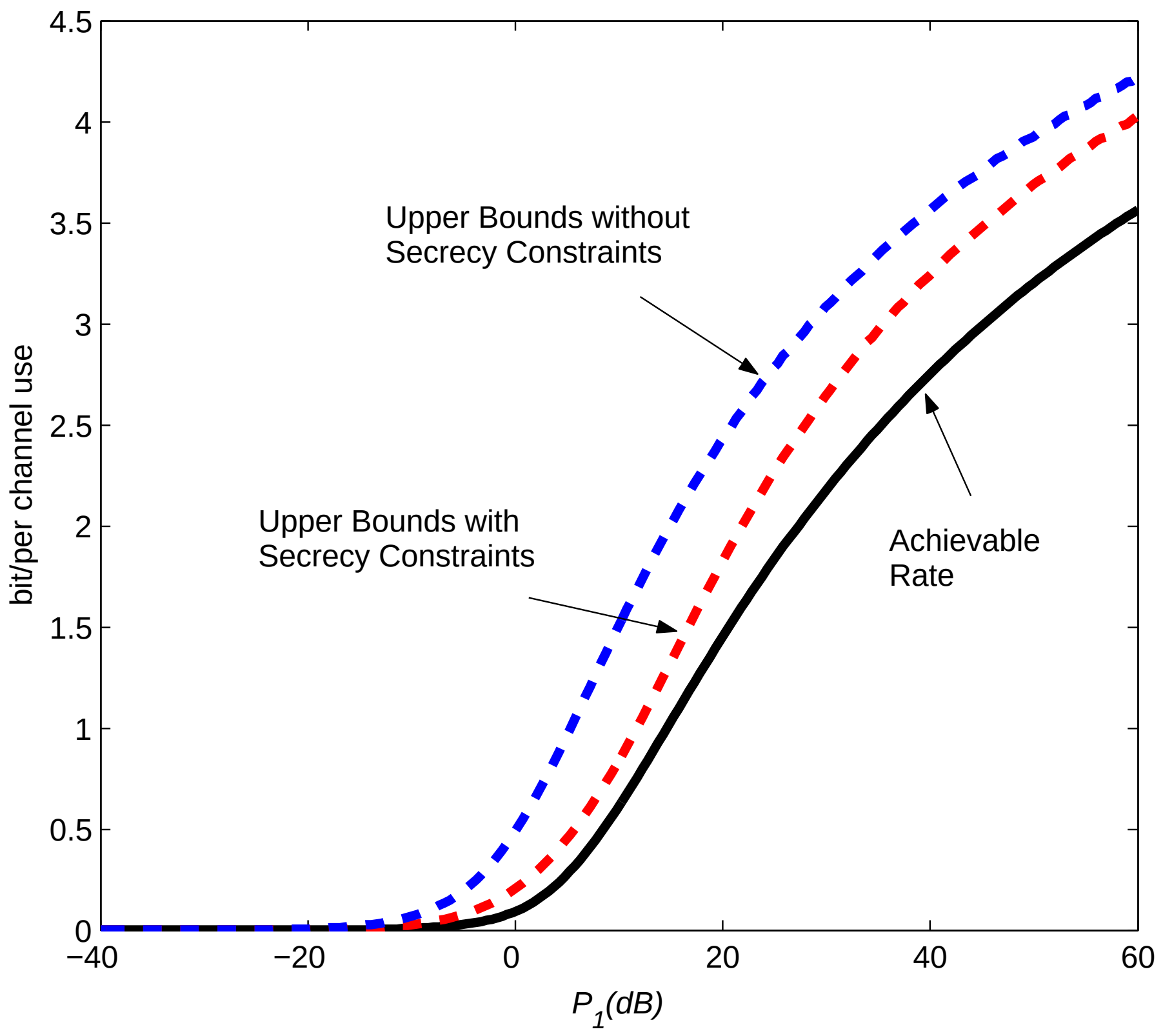

Fig. 10. Secrecy Rate, $P_r = 40$dB, $P_2 = 0.25P_1$, optimized over $\alpha$

*Theorem 3:* Consider a relay network with conditional distribution $p(Y, Y_r|X, X_r)$, with $X$, $X_r$ being the input from the source and the relay respectively, and $Y_r$,$Y$ being the signals received by the relay and the destination respectively. For the distribution

$$p(X)p(X_r)p(Y, Y_r|X, X_r)p(\hat{Y}_r|Y_r, X_r) \tag{60}$$

the following range of rates $R$ is achievable.

$$0 \leq R < [I\left(X; Y\hat{Y}_r|X_r\right) - I\left(X; Y_r|X_r\right)]^+ \tag{61}$$

with

$$I(X_r; Y) > I(\hat{Y}_r; Y_r|YX_r) \tag{62}$$

Theorem 3 follows from the achievable equivocation region given in [11, Theorem 1] by simply considering rates $R$ that equal the equivocation rate $R_e$. The proof of Theorem 3 is given in

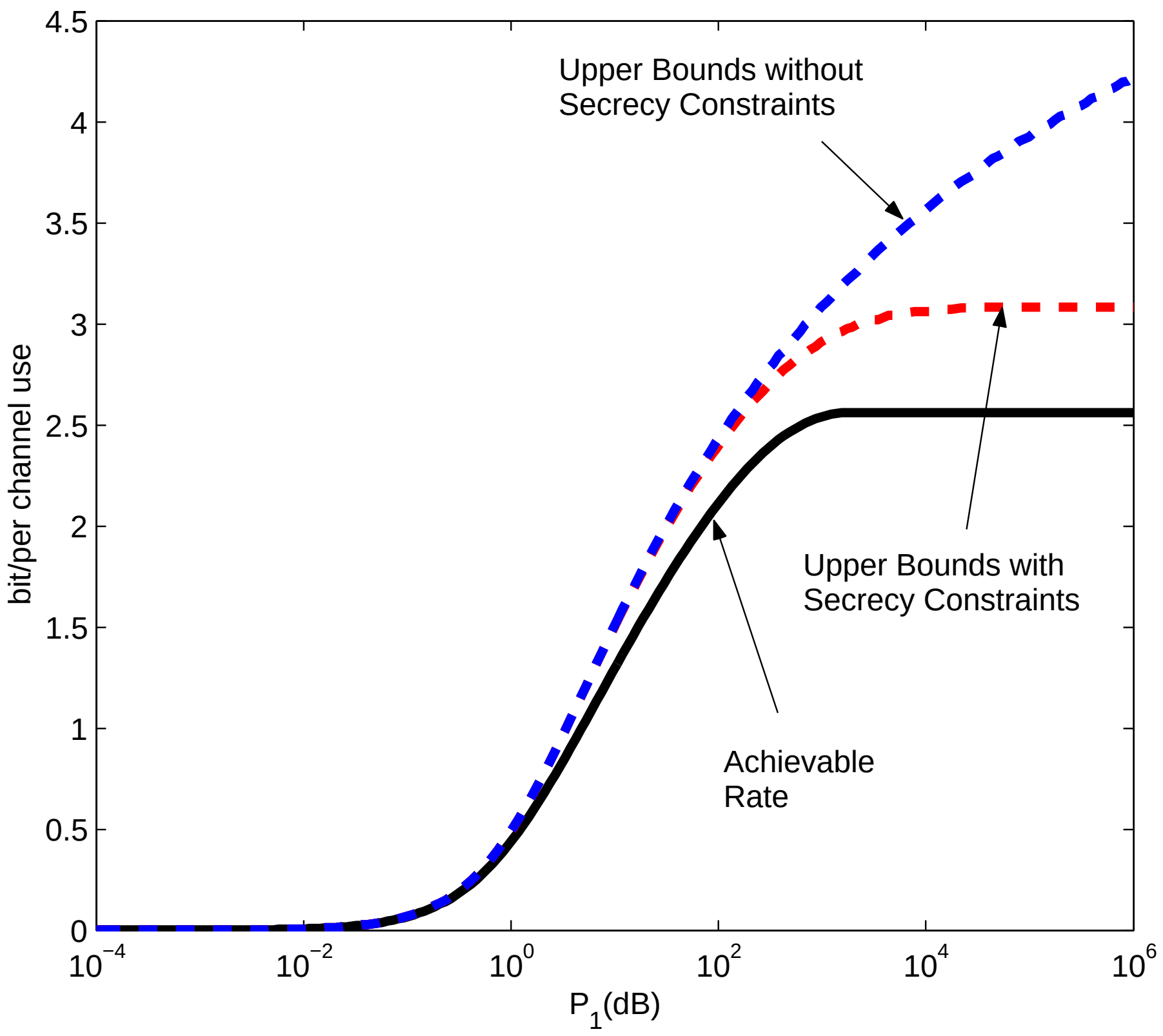

Fig. 11. Secrecy Rate, $P_r = 40$dB, $P_2 = 30$dB, optimized over $\alpha$, power control at the source node enabled

[11]. The outline of the achievable scheme is as follows: The relay does compress-and-forward as described in [25]. Therefore, as in [25], $X_r$ is independent from $X$ in the input distribution expression (60). The same decoder in [25] is used at the destination. The same codebook as [25] is used at the source node. However, instead of mapping the message to the codeword deterministically as in [25], a stochastic encoder is used at the source node. In this encoder, the codewords are randomly binned into several bins. The size of each bin is $2^{NI(X;Y_r|X_r)}$ where $N$ is the total number of channel uses. The message $W$ determines which bin to use by the encoder. The actual transmitted codeword is then randomly chosen from the bin according to a uniform distribution. This randomness serves to confuse the eavesdropper at the relay node at the cost of the rate as shown by the term $-I\left(X;Y_r|X_r\right)$ in (61).

We next extend this result by considering a relay channel with a jammer defined by

$$p(Y_r, Y|X, X_2, X_r) \tag{63}$$

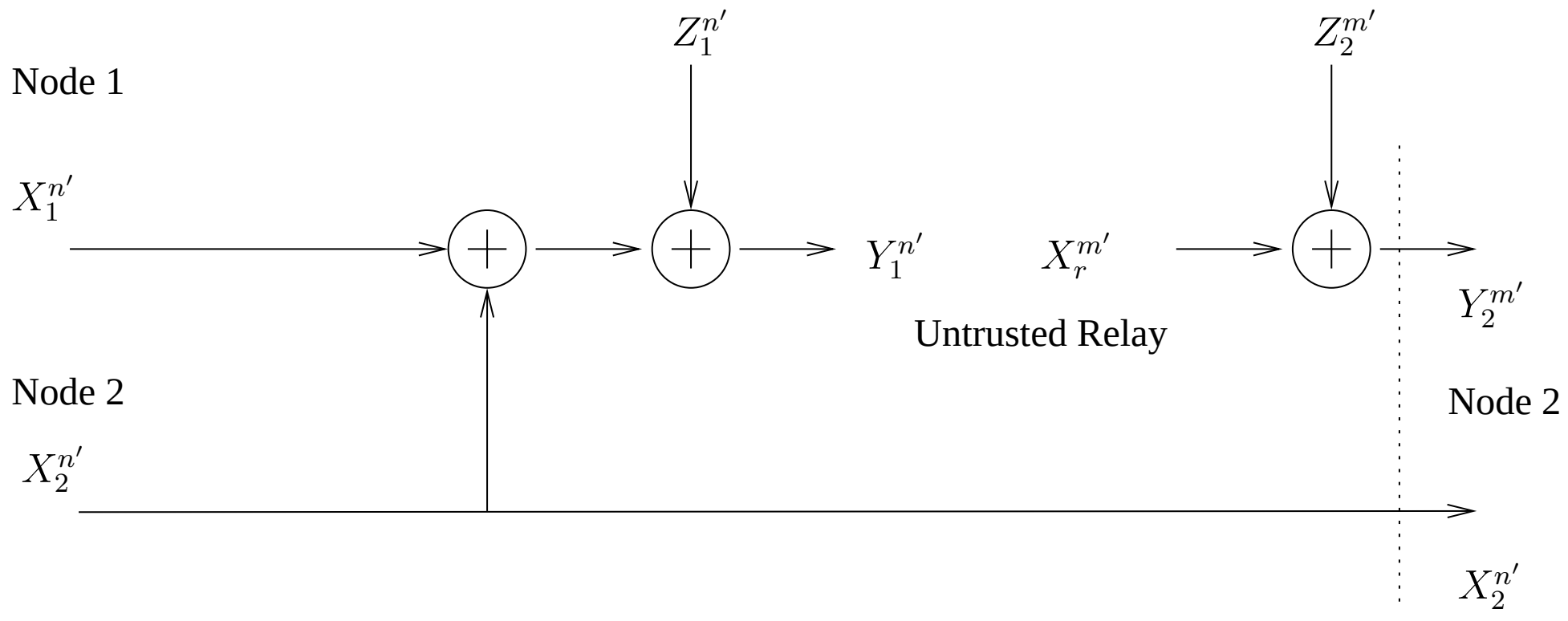

Fig. 12. Equivalent Channel Model

where $X_2$ is the signal transmitted by the jammer and the notation $Y, Y_r, X, X_2$ follows the definition above. Then, if the jammer transmits an i.i.d. signal according to distribution $p(X_2)$ and $p(X, X_2, X_r) = p(X)p(X_2)p(X_r)$, the induced channel $p(Y_r, Y|X, X_r)$ is given below

$$p\left(Y_r, Y|X, X_r\right) = \sum_{X_2} p\left(X_2\right) p\left(Y_r, Y|X, X_2, X_r\right) \tag{64}$$

and it is also a memoryless relay channel. Hence, we can use Theorem 3 and obtain the following corollary:

*Corollary 1:* The following secrecy rate is achievable:

$$0 \le R \le \max_{p(X)p(X_2)p(X_r)p(Y,Y_r|X,X_2,X_r)p(\hat{Y}_r|Y_r,X_r)} \left[ I\left(X; Y\hat{Y}_r|X_r\right) - I\left(X; Y_r|X_r\right)\right]^+ \tag{65}$$

with

$$I(X_r; Y) > I(\hat{Y}_r; Y_r|YX_r) \tag{66}$$

We next reformulate our channel in figure 1 in a way such that Corollary 1 can be applied. This is shown in Figure 12. Here we can draw the jammer and the receiver separately, since the jammer does not use the signal received in the past to compute the jamming signal. The $m'$ and $n'$ are the parameters of the schedule described in Section III. We then observe Figure 12 can be viewed as a three node relay network with a jammer, defined as follows:

$$p\left(\mathrm{Y}, \mathrm{Y}_r|\mathrm{X}, \mathrm{X}_2, \mathrm{X}_r\right) \tag{67}$$

where

$$\mathrm{Y} = \left\{Y_2^{m'}, X_2^{n'}\right\} \tag{68}$$

$$\mathrm{Y}_r = Y_1^{n'}, \quad \mathrm{X} = X_1^{n'} \tag{69}$$

$$\mathrm{X}_r = X_r^{m'}, \quad \mathrm{X}_2 = X_2^{n'} \tag{70}$$

The input distributions to this vector input channel are chosen as

$$p(X_1^{n'}, X_2^{n'}, X_r^{m'}) = p(X_1^{n'})p(X_2^{n'})p(X_r^{m'}) \tag{71}$$

where $p(X_1^{n'})$, $p(X_2^{n'})$ and $p(X_r^{m'})$ are given below:

1) Let $X_1^{n'} \sim \mathcal{N}(0, P_1'\mathbf{I}_{n'\times n'})$, where $P_1'$ is the average power consumption of node $1$ during the periods of phase one. Hence $0 < P_1' < P_1$.
2) Let the auxiliary random variable in compress-and-forward $\hat{\mathrm{Y}}_\mathrm{r}$ be $\hat{Y}_1^{n'}$. Let $\hat{Y}_1^{n'} = Y_1^{n'} + Z_Q^{n'}$, where $Z_Q^{n'} \sim \mathcal{N}(0, \sigma_c^2\mathbf{I}_{n'\times n'})$.
3) Let $X_r^{m'} \sim \mathcal{N}(0, P_r\mathbf{I}_{m'\times m'})$ and $X_2^{n'} \sim \mathcal{N}(0, P_2\mathbf{I}_{n'\times n'})$.

where $\mathbf{I}_{n'\times n'}$ denotes an $n' \times n'$ identity matrix. With 1)-3), we have

$$I\left(\mathrm{X}; \mathrm{Y}\hat{\mathrm{Y}}_\mathrm{r}|\mathrm{X}_\mathrm{r}\right) \tag{72}$$

$$=I\left(X_1^{n'}; Y_2^{m'}X_2^{n'}\hat{Y}_1^{n'}|X_r^{m'}\right) \tag{73}$$

$$=I\left(X_1^{n'}; Y_2^{m'}\hat{Y}_1^{n'}|X_r^{m'}X_2^{n'}\right) + I\left(X_1^{n'}; X_2^{n'}|X_r^{m'}\right) \tag{74}$$

From (71), $I\left(X_1^{n'}; X_2^{n'}|X_r^{m'}\right) = 0$. Therefore (74) equals

$$I\left(X_1^{n'}; Y_2^{m'}\hat{Y}_1^{n'}|X_r^{m'}X_2^{n'}\right) \tag{75}$$

$$=I\left(X_1^{n'}; \hat{Y}_1^{n'}|X_r^{m'}Y_2^{m'}X_2^{n'}\right) + I\left(X_1^{n'}; Y_2^{m'}|X_r^{m'}X_2^{n'}\right) \tag{76}$$

$$=I\left(X_1^{n'}; \hat{Y}_1^{n'}|X_r^{m'}Y_2^{m'}X_2^{n'}\right) + I\left(X_1^{n'}; Z_2^{m'}|X_r^{m'}X_2^{n'}\right) \tag{77}$$

$$=I\left(X_1^{n'}; \hat{Y}_1^{n'}|X_r^{m'}Y_2^{m'}X_2^{n'}\right) \tag{78}$$

(78) equals:

$$I\left(X_1^{n'}; Y_1^{n'} + Z_Q^{n'}|X_r^{m'}Y_2^{m'}X_2^{n'}\right) \tag{79}$$

$$=I(X_1^{n'}; X_1^{n'} + X_2^{n'} + Z_1^{n'} + Z_Q^{n'}|X_r^{m'}, X_r^{m'} + Z_2^{m'}, X_2^{n'}) \tag{80}$$

$$=I\left(X_1^{n'}; X_1^{n'} + X_2^{n'} + Z_1^{n'} + Z_Q^{n'}|X_2^{n'}\right) \tag{81}$$

$$=I\left(X_1^{n'};X_1^{n'}+Z_1^{n'}+Z_Q^{n'}\right) \tag{82}$$

$$=n'C\left(\frac{P_1'}{1+\sigma_c^2}\right) \tag{83}$$

and

$$I\left(\mathrm{X};\mathrm{Y_r}|\mathrm{X_r}\right) \tag{84}$$

$$=I\left(X_1^{n'};Y_1^{n'}|X_r^{m'}\right) \tag{85}$$

$$=I\left(X_1^{n'};X_1^{n'}+X_2^{n'}+Z_1^{n'}|X_r^{m'}\right) \tag{86}$$

$$=I\left(X_1^{n'};X_1^{n'}+X_2^{n'}+Z_1^{n'}\right) \tag{87}$$

$$=n'C\left(\frac{P_1'}{1+P_2}\right) \tag{88}$$

and

$$I\left(\mathrm{X_r};\mathrm{Y}\right) \tag{89}$$

$$=I\left(X_r^{m'};Y_2^{m'},X_2^{n'}\right) \tag{90}$$

$$=I\left(X_r^{m'};Y_2^{m'}\right) \tag{91}$$

$$=m'C\left(P_r\right) \tag{92}$$

and

$$I\left(\hat{\mathrm{Y}}_r;\mathrm{Y}_r|\mathrm{Y},\mathrm{X_r}\right) \tag{93}$$

$$=I\left(\hat{Y}_1^{n'};Y_1^{n'}|Y_2^{m'}X_2^{n'}X_r^{m'}\right) \tag{94}$$

$$=I\left(Y_1^{n'}+Z_Q^{n'};Y_1^{n'}|X_2^{n'}X_r^{m'}Z_2^{m'}\right) \tag{95}$$

$$=I(X_1^{n'}+X_2^{n'}+Z_1^{n'}+Z_Q^{n'};X_1^{n'}+X_2^{n'}+Z_1^{n'}|X_2^{n'},X_r^{m'}) \tag{96}$$

$$=I(X_1^{n'}+X_2^{n'}+Z_1^{n'}+Z_Q^{n'};X_1^{n'}+X_2^{n'}+Z_1^{n'}|X_2^{n'}) \tag{97}$$

$$=I\left(X_1^{n'}+Z_1^{n'}+Z_Q^{n'};X_1^{n'}+Z_1^{n'}\right) \tag{98}$$

$$=n'C\left(\frac{P_1'+1}{\sigma_c^2}\right) \tag{99}$$

In (75) (81) (87) (91) (97) (98), we use (71) repeatedly, which says that with compress-and-forward, the input distribution are chosen such that $X_1^{n'},X_2^{n'},X_r^{m'}$ are independent.

Substituting the values of $I\left(\mathrm{X};\mathrm{Y}\hat{\mathrm{Y}}_{\mathrm{r}}|\mathrm{X}_{\mathrm{r}}\right)$, $I\left(\mathrm{X};\mathrm{Y}_{\mathrm{r}}|\mathrm{X}_{\mathrm{r}}\right)$, $I\left(\mathrm{X}_{\mathrm{r}};\mathrm{Y}\right)$ and $I\left(\hat{\mathrm{Y}}_{r};\mathrm{Y}_{r}|\mathrm{Y},\mathrm{X}_{\mathrm{r}}\right)$ into Corollary 1, dividing both sides by $m'+n'$, and taking the limit $m'+n'\rightarrow\infty$, we proved the theorem.